\documentclass[english]{cidarticle}
\usepackage{blindtext} 

\usepackage{booktabs}
\usepackage{listings}
\usepackage{csquotes}
\usepackage{url}
\usepackage{tikz}
\usepackage{wrapfig}

\newcommand{\out}[1]{\iffalse{#1}\fi}
\def\defemph#1{\textbf{#1}}

\usepackage{tcolorbox,xspace}

\definecolor{dunkelblau}{cmyk}{1,0.5,0.1,0.375}
\definecolor{hellgrau}{cmyk}{0,0,0,0.1}
\definecolor{dunkelrot}{cmyk}{0.1,1,1,0.375}

\DeclareTColorBox{Task}{ O{dunkelblau} O{hellgrau} O{} }{
	title=Task,
	colframe=#1,
	colback=#2,
}

\newcommand{\VOLLKORN}{\texttt{VoLl-KOrN}\xspace}

\begin{document}

\title[Potential Answer Classes]{The Potential of Answer Classes in Large-scale Written Computer-Science Exams -- Vol. 2}

\author[1]{Dominic Lohr}{dominic.lohr@fau.de}{0000-0002-6330-2327}
\author[1]{Marc Berges}{marc.berges@fau.de}{0000-0002-9982-547X}
\author[2]{Michael Kohlhase}{michael.kohlhase@fau.de}{0000-0002-9859-6337}
\author[2]{Florian Rabe}{florian.rabe@fau.de}{0000-0003-3040-3655}
\affil[1]{Friedrich-Alexander-Universität Erlangen-Nürnberg\\Didaktik der Informatik\\Martensstraße 3\\91058 Erlangen\\Deutschland}
\affil[2]{Friedrich-Alexander-Universität Erlangen-Nürnberg\\Wissensrepräsentation und -verarbeitung\\Martensstraße 3\\91058 Erlangen\\Deutschland}

\maketitle

\begin{abstract}
Students' answers to tasks provide a valuable source of information in teaching as they result from applying cognitive processes to a learning content addressed in the task. Due to steadily increasing course sizes, analyzing student answers is frequently the only means of obtaining evidence about student performance. However, in many cases, resources are limited, and when evaluating exams, the focus is solely on identifying correct or incorrect answers. This overlooks the value of analyzing incorrect answers, which can help improve teaching strategies or identify misconceptions to be addressed in the next cohort.  

In teacher training for secondary education, \emph{assessment guidelines} are mandatory for every exam, including anticipated errors and misconceptions. We applied this concept to a university exam with 462 students and 41 tasks. For each task, the instructors developed \emph{answer classes} -- classes of expected responses, to which student answers were mapped during the exam correction process. The experiment resulted in a shift in mindset among the tutors and instructors responsible for the course: after initially having great reservations about whether the significant additional effort would yield an appropriate benefit, the procedure was subsequently found to be extremely valuable. 

The concept presented, and the experience gained from the experiment were cast into a system with which it is possible to correct paper-based exams on the basis of answer classes. This updated version of the paper provides an overview and new potential in the course of using the digital version of the approach.
\end{abstract}

\section{Introduction}
\label{sec:introduction}

A reliable and fair assessment of exam assignments is a fundamental requirement for professional teaching.
Especially in large-scale written exams, resources are limited, and high-quality assessment of submissions is challenging for educators. Inconsistencies in correction are possible when multiple people correct the same assignment -- especially if no high-quality assessment guidelines are available and where the correcting educator has a high degree of decision-making power. In addition, the feedback provided to students is very basic at best and is often only provided in scoring (points).

To support correction processes to make it fairer and more objective, in secondary and higher education, \emph{marking schemes} respectively \emph{rubrics} \cite{Ahmed.2011} are mandatory when creating assignments -- especially exams. The advantage is that it reduces required corrections and provides more objective task evaluation options. In addition to the model solution, \emph{rubrics} contain alternative solutions and a list of possible incorrect answers. This led us to the concept of \emph{answer classes} -- groups of answers that share the same underlying idea. The intelligent clustering of possible solutions into answer classes holds several great potentials for modern teaching -- especially when dealing with so many learners that individual face-to-face support is impossible.

To investigate the potential of answer classes in large-scale written computer science exams, we developed a process model that we evaluated in the first iteration of an exam for an Artificial Intelligence major program that 462 students took.
Following the process model, 192 answer classes were developed for 41 tasks.
In the sequel, we present a detailed analysis of five tasks aimed at identifying the specific (possible) causes of incorrect answers and feeding those back into next year's teaching process. The results demonstrate the efficacy of answer classes in creating rubrics for large courses and highlight their significant potential for enhancing instructional materials, improving the fairness of correction, and developing adaptive feedback.

A completely unexpected consequence was that the experiment resulted in a paradigm shift among the tutors and instructors responsible for the course, who -- in a university setting -- were not accustomed to rubric practices in secondary education. Despite initial reservations regarding the additional effort required, the procedure was ultimately deemed valuable and worthy of undertaking in every future exam.

This rethink ultimately led to efforts to digitize this process. Based on the experience gained from the first manual iteration, a system was developed that makes it possible to correct scanned paper exams based on answer classes and provide feedback based on these classes.

\textbf{Note}: This paper is a revised version of the 2023 publication \cite{lohrPotentialAnswerClasses2023}. In the previous version, the system just mentioned was future work and is now -- one year later -- already put to practice and has already been used in 10 exams with 10 to 300 students each. The revised paper has been updated to reflect new experiences and potential.

\section{Related Work}\label{sec:related-work}

Classifying answers in examinations or tasks, in general, is an essential but tedious part of correcting and rating written exams. It is no surprise that the increasing power of computers has led to efforts to automate that.

\paragraph{Auto-grading systems and feedback generation} often refer to a set of rules defined by the author of the task. For instance, this is often done in programming using static and dynamic code analysis, as seen in the system \texttt{JACK}, which uses a query language in XML to specify test cases and code structures relevant for grading \cite{Striewe2016}. \texttt{JACK} handles diverse tasks like UML modeling and fill-in-the-gap assignments. Many e-assessment systems detect errors for complex programming tasks by executing test cases and analyzing source code \cite{Ihantola.2010, Strickroth.2022}. Automated grading matches human performance \cite{Gaudencio.2014}, with fine-grained feedback being effective \cite{Falkner.2014}. In contrast, Course Master by Higgins et al. integrates marking schemes, separating exercises and grading logic as Java classes, enabling parameterized tasks, marks (grades), and feedback \cite{Higgins.2002}. Lawrence et al. explore performance-based grading, crucial for e-learning and remote exams \cite{Lawrence.2023}.

\paragraph{Answer Clustering/Classification} A different approach to classifying student answers is conducted by Zehetmeier et al. \cite{Zehetmeier.2015}. They analyzed student submissions to identify and cluster errors and quantified significant groups to uncover the causes of errors. The researchers also observed the students while programming, interviewed them about their errors at crucial stages, and used the results to create teaching materials and tasks to prevent similar misconceptions. In their system \texttt{EvalSeer}, Nabil et al. use machine learning for classifying answers and propose syntax fixes with an appropriate accuracy \cite{Nabil.2021}. While many studies focus on programming tasks due to their complexity, there are other complex tasks in computer science, such as marking up UML database diagrams \cite{Foss.2022}.

\section{Theoretical Background}
\label{sec:theoretical-background}

\begin{wrapfigure}r {0.55\textwidth}\centering          
    \includegraphics[width=6cm]{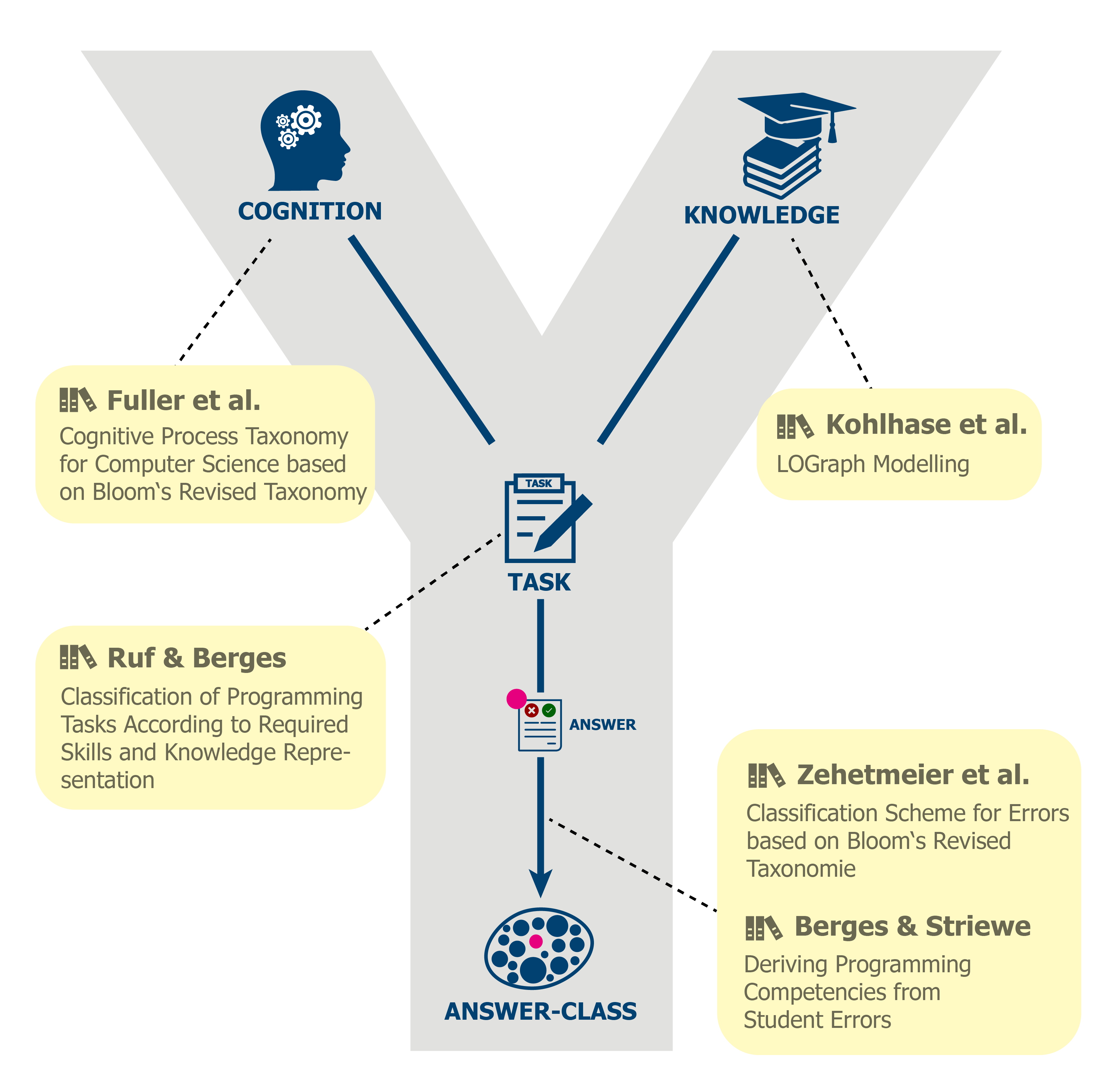}
    \caption{The Y-model}\label{fig:overview}\vspace*{-.5em}
\end{wrapfigure}

The concept of answer classes (AC) is one of four components of the Y-Model \cite{LohBerKoh:ymodel23} -- a model for formalizing tasks in Computer Science (see \autoref{fig:overview}). The model forms a basis for a competence-oriented integration of tasks in adaptive learning systems. 

The central component of the Y-Model is the task itself, represented by its description (and representation). The upper part illustrates the interaction of knowledge elements with cognitive processes.
Knowledge elements and their interdependencies are modeled using \emph{knowledge/learning object graphs} \cite{KohKoh:skmfe08}. A learning taxonomy like the one by Fuller et al.\cite{Fuller.2007} or Bloom's revised  \cite{Anderson.2009} can be used to model cognitive processes. For a detailed overview of the task model's components and underlying theories, see \cite{LohBerKoh:ymodel23}.

\paragraph{Answer classes (ACs)} (see lower part of the Y-model) can be understood as a set-theoretic propositional form. 
Each answer class includes an identifier $AC_x$ -- unique to the task -- and a detailed answer class description \emph{ACD}, outlining the criteria for assigning a given answer to this class. \emph{AC}s are notably not free of intersections so that answers can be assigned to several different \emph{AC}s. For the \emph{AC} to be reliably annotated by any educator, the \emph{ACD} must be (1) unambiguous (free of interpretation) and (2) objectively observable. Therefore, criteria that refer to presumed causes of errors, such as ``sloppy work'' or ``lack of understanding'' are not permitted, as these are merely hypotheses without further information about the learner. The criteria must be formulated objectively and free of interpretation so that a reliable classification by various independent graders -- and thus a fairer correction process -- is possible.

A general distinction is made between context-independent and task-specific answer classes. The former are applicable across various tasks, while the latter are related to a specific type of task or topic/subject. Examples of context-independent answer classes are $\text{AC}_1 = \{R \,|\, R\text{ is empty} \}$, which contains all answers $R$ that have not been processed, and $\text{AC}_2 = \{R \,|\, R\text{ is crossed out} \}$, which contains answers that have been written but crossed out. Regarding scoring, these two answer classes may have the same effect. However, from a didactic perspective, educators may distinguish whether a task was not worked on or was attempted. Task-specific answer classes only "live" within a specific task, a specific type of task, or a specific domain. 

In the context of adaptive learning systems, the concept of answer classes enables a selection process of tasks based on prior knowledge and competency goals, as well as on answers given by students from previous tasks. In addition, specific error patterns, such as those proposed by Zehetmeier et al. ~\cite{Zehetmeier.2015}, or Berges et al. \cite{Berges.2016}, can be integrated into answer classes and, for instance, infer possible causes of errors by referring to a corresponding learner model. In addition, answer classes can act as keys for specifying feedback or hints on how to proceed. 

\paragraph{Guidelines for assessment and evaluation} are common in teaching. Still, there is a wide range of terminologies utilized to describe them, including \emph{Grading scheme}, \emph{Marking Scheme}, \emph{Rubric}, and \emph{Erwartungshorizont}. These guidelines are used to evaluate student work, but they differ in some essential aspects:

\emph{Grading Schemes} outline the percentage or numerical value assigned to different levels of achievement \cite{Lofgren.2007}, while \emph{Marking Schemes} provide a breakdown of the criteria used to evaluate the work and the weight assigned to each criterion \cite{Ahmed.2011}. \emph{Rubrics} \cite{Luft1999} \cite{Stevens_Levi_2005}, on the other hand, provide a detailed description of the criteria for each level of achievement, along with examples of what work at each level might look like. They are widely used in American education, from K-12 to higher education institutions. They provide clear and consistent student feedback, guide instruction and assessment, and help ensure fairness and objectivity in grading. Rubrics can be developed for various assignments and assessment types and customized to fit the specific needs of different courses and instructors. \emph{Erwartungshorizont} is a German term that refers to a set of expectations or guidelines used to evaluate student work, similar to a grading scheme or marking scheme \cite{KotterMathes.2020b}. Overall, while all of these terms serve a similar purpose, they differ in their level of detail and the specific information they provide to educators and students. Since the concept of rubrics best fits the potential of answer classes explored in this paper, we decided to use this term.

\section{The Potential of answer classes}
\label{sec:potential}

Developing and implementing answer classes involves additional effort for educators, but we claim that the potential benefits are significant. To evaluate the potential of ACs, we begin by establishing criteria derived from the advantages of rubrics in the literature (e.g., \cite{Wolf.2007}, \cite{Reddy.2010}):

\begin{description}
\item[ST] \emph{Save time} Conventional correction process, involving marginal notes and closing comments, can be time-consuming. Answer classes reduce the need for these annotations, potentially making the correction process more efficient.

\item[BF] \emph{Better feedback}  In most cases, the lack of time during corrections leads to the omission of valuable feedback. However, developing feedback for each answer class must only be done once (or for their combinations) and can be effortlessly provided as a table.

\item[MOC] \emph{More objective correction} Multiple educators correct the same task in large courses, leading to disparate results due to varying personal interpretations. Answer classes offer a set of precisely defined, objectively observable criteria, enhancing the reliability of the correction process.

\item[BPE] \emph{Better pre-estimation} Addressing potential student errors in advance helps to ensure that exam tasks are appropriate and error-free.

\item[IIT] \emph{Incentives to improve teaching} The annotation of answers with ACs allows for a higher quality evaluation of exam results, providing valuable insights for teaching improvement. Prevalent error classes can indicate particular areas in the teaching material that must be revised to avoid misconceptions.
\end{description}

\section{Methodology}
\label{sec:methodology}

\begin{figure}\centering
    \includegraphics[width=\linewidth]{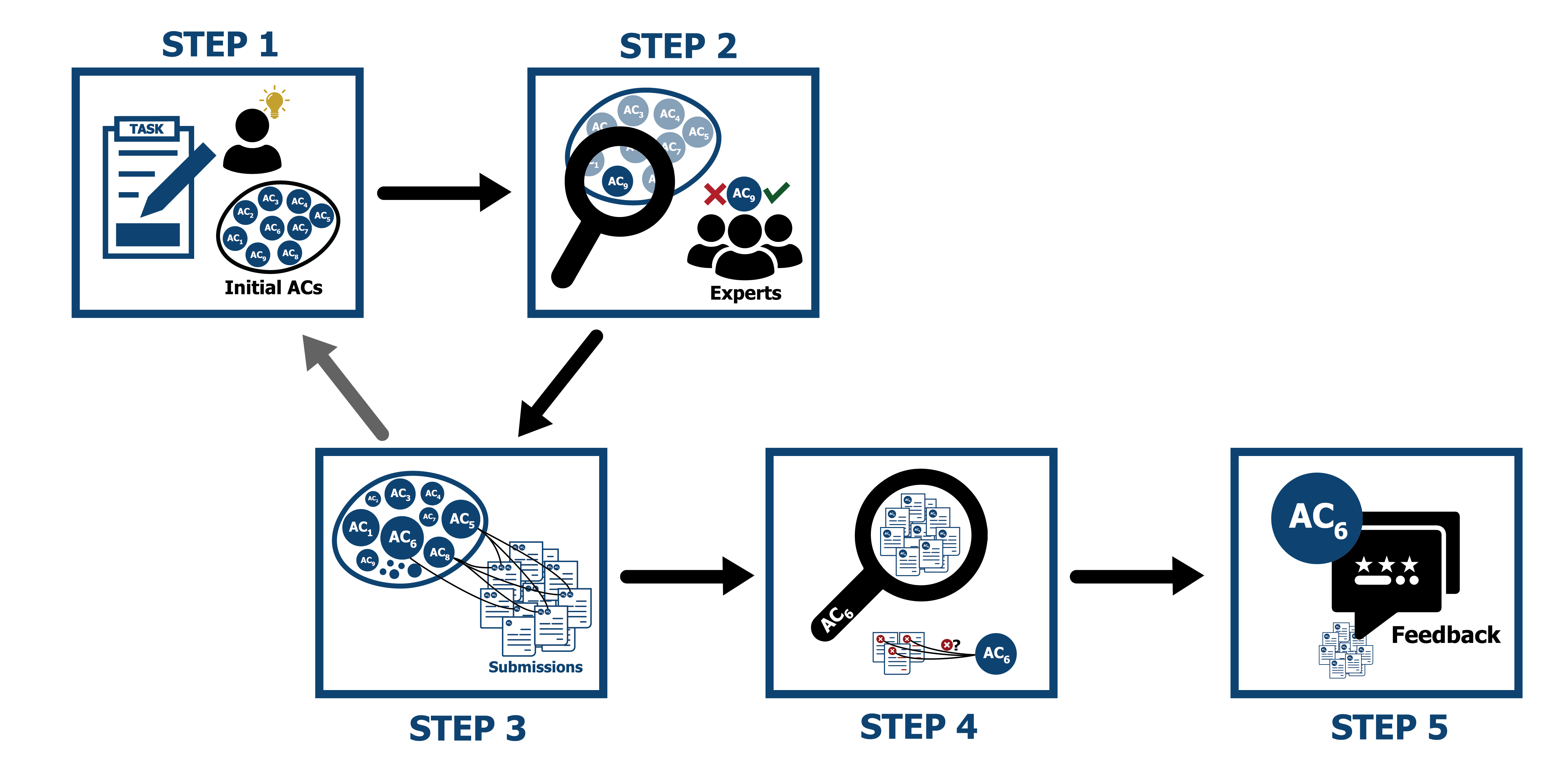}
    \caption{Iterative Process of AC-Creation}\label{fig:process}
\end{figure}

To investigate the potential categories stated in Section \ref{sec:potential}, we devised an iterative process model to develop and apply the concept of answer classes systematically (see \Cref{fig:process}).

Upon creation of the exam, the instructors were introduced to the concept of answer classes as detailed in Section \ref{sec:theoretical-background}. To begin with, a preliminary set of answer classes was created for each task based on prior experience with similar tasks (STEP 1).

Subsequently, we discussed the initial answer classes for each task in a plenary session (STEP 2). For this purpose, the instructors presented the tasks and the first draft of the set of answer classes to the group. During this session, all stakeholders checked that the answer classes were objectively observable and free of subjectivity or ambiguity. 

\begin{figure}\centering
    \includegraphics[width=\linewidth]{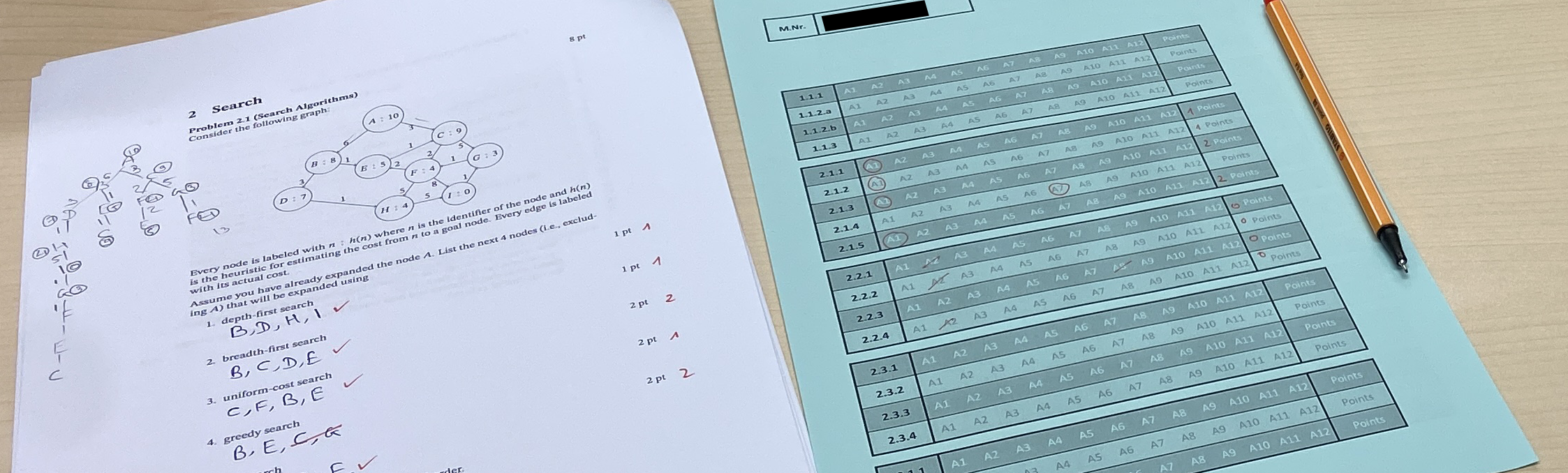}
    \caption{Exam Sheet and AC Mapping Form}\label{fig:mapping-list}
\end{figure}

Considering that the exam was administered as a paper-based test, we developed an \defemph{AC mapping form} to facilitate the mapping process of students' submissions to corresponding answer classes (STEP 3). The list contained -- besides the students' IDs -- a table with all the identifiers of the ACs for each task.
(see \autoref{fig:mapping-list}).

To ensure the correct mapping of answer classes by the correctors, we added a brief explanation of each answer class to the model solution. The AC mapping form also contained placeholders for (potential) new answer classes that were not included during STEP 1 and identified during the correction process. Candidates for such new answer classes are usually (incorrect) answers that occur very frequently but cannot yet be assigned to an existing answer class.

Based on a descriptive analysis of the results in STEP 3, we selected ACs with frequent occurrences. Further, we scrutinized them in STEP 4, where we analyzed sample exams of the respective AC, intending to identify possible underlying factors that led the student to that answer. 

The outcomes of STEP 4 informed the development of feedback for each answer class. The development of AC identifiers for automated annotation of submissions with their respective AC can be incorporated in this step but was left as future work.

Assuming exam task types are reused in subsequent exams, the process described is to be understood as iterative: the revised set of answer classes after STEP 3 serves as the initial set in STEP 1 of the subsequent iteration.

\section{Results}
\label{sec:results}

The procedures outlined in Section \ref{sec:methodology} were initially tested in an exam for the course \emph{Artificial Intelligence 1}. Of the 502 registered students, 462 participated in the written exam (90 minutes, conducted in presence). The majority of the students were pursuing their master's degree in Data Science (225), Artificial Intelligence (118), and Computer Science (69). The exam consisted of 12 problems, encompassing 41 distinct tasks. STEP 1 to 3 were done for all 41 tasks. A total of four common and 192 initial answer classes were developed. Since steps 4 and 5 were more time-intensive, we selected one problem (containing five tasks) and executed these steps. The selected problem focused on searching in a directed graph with edge costs and node heuristics. The concrete objective was to apply five different search algorithms (e.g., $A^*$-Search, Greedy-Search, ...) starting from an already expanded node and to give the subsequent four visited nodes (for a complete problem description, see \autoref{fig:task-description}). This section describes the application of steps 1-5 to this problem.

\begin{figure}
\begin{Task}

\textbf{Problem 2.1 (Search Algorithms)} \hfill 8 pt\\
Consider the following graph:

\begin{center}
  \begin{tikzpicture}[yscale=1.5]
  \node[circle,draw] (0) at (0,0) {$A:10$};
  \node[circle,draw] (1) at (-2,-1) {$B:8$};
  \node[circle,draw] (2) at (2,-1) {$C:9$};
  \node[circle,draw] (3) at (-3,-2) {$D:7$};
  \node[circle,draw] (4) at (-0.5,-1.5) {$E:5$};
  \node[circle,draw] (5) at (1,-2) {$F:4$};
  \node[circle,draw] (6) at (3,-2) {$G:3$};
  \node[circle,draw] (7) at (0,-3) {$H:4$};
  \node[circle,draw] (8) at (2,-3) {$I:0$};  
  \draw[->] (0) --node[left] {6} (1);
  \draw[->] (0) --node[left] {3} (2);
  \draw[->] (1) --node[left] {3} (3);
  \draw[->] (1) --node[above] {1} (4);
  \draw[->] (4) --node[above] {1} (2);
  \draw[->] (5) --node[above] {2} (4);
  \draw[->] (7) --node[above] {5} (8);
  \draw[->] (3) --node[above] {1} (7);
  \draw[->] (2) --node[left] {2} (5);
  \draw[->] (2) --node[left] {5} (6);
  \draw[->] (5) --node[left] {5} (7);
  \draw[->] (5) --node[left] {8} (8);
  \draw[->] (8) --node[left] {1} (6);
  \draw[->] (6) --node[above] {1} (5);
  \end{tikzpicture}
\end{center}

Every node is labeled with $n : h(n)$ where $n$ is the identifier of the node and $h(n)$ is the heuristic for estimating the cost from $n$ to a goal node. Every edge is labeled with its actual cost.

Assume you have already expanded the node $A$. List the next four nodes (i.e., excluding $A$) that will be expanded using.

\begin{enumerate}
    \item depth-first search \hfill 1 pt.
    \item breadth-first search \hfill 1 pt.
    \item uniform-cost search \hfill 2 pt.
    \item greedy search \hfill 2 pt.
    \item A*-search \hfill 2 pt.
\end{enumerate}

If there is a tie, break it using alphabetical order.
\end{Task}
\caption{The analyzed task consisting of 5 subtasks}
\label{fig:task-description}
\end{figure}

\begin{table*}[h]
\centering
\begin{tabular}[]{l|l}
    ID & Answer Class\\ 
    \hline
    $\text{AC}_1$ & $\{R \,|\, R\text{ is empty} \}$ \\
    $\text{AC}_2$ & $\{R \,|\, R\text{ is crossed out} \}$ \\
    $\text{AC}_3$ & $\{R \,|\, R\text{ is fully correct} \}$ \\
    $\text{AC}_4$ & $\{R \,|\, \text{no suitable AC known} \}$ \\
    $\text{AC}_5$ & $\{R \,|\, \text{the first node of R is correct}  \}$ \\
    $\text{AC}_6$ & $\{R \,|\, \text{the second node of R is correct } \}$ \\
    $\text{AC}_7$ & $\{R \,|\, \text{the third node of R is correct} \}$ \\
    $\text{AC}_8$ & $\{R \,|\, \text{the fourth node of R is correct} \}$ \\
\end{tabular}
\caption{List of Initial ACs for Problem 2.1 (Step 2)}
\label{table:initial-ac}
\end{table*}

Table \ref{table:initial-ac} shows the initial set of ACs for problem 2.1. $AC_1$ to $AC_4$ are the context-independent classes that are the same for all tasks.   
Since the tasks differ only in the search algorithm to be applied, the initially conceived task-specific answer classes $AC_5$ to $AC_8$ were also the same. 

It turned out that these initial ACs were not very helpful for further analysis of the answers because the information was limited. They were useful only in correction since it was possible to determine the exact score from the combination of ACs assigned to an answer. However, the feedback that could be generated from this was only elementary. 
For this reason, in addition to the AC-mapping process (Step 3), various answers were recorded in a table via frequency analysis to find an improved set of new answer classes for these tasks.  

\begin{figure}\centering
    \includegraphics[width=\linewidth]{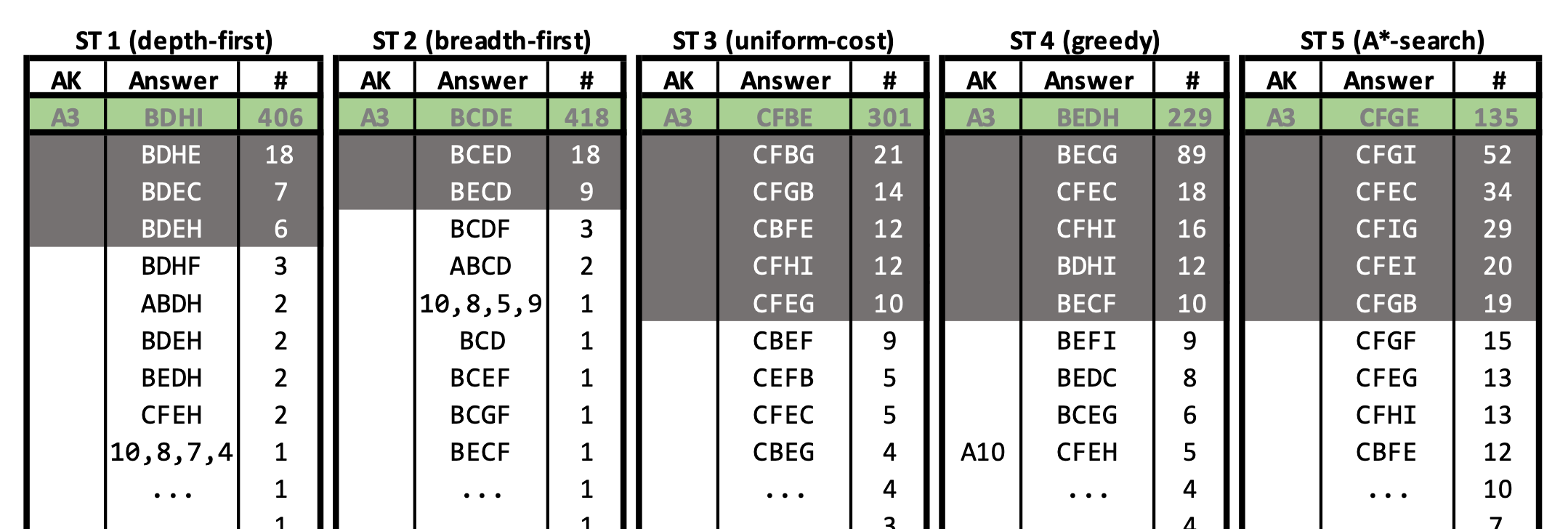}
    \caption{Frequency analysis of given answers to the tasks}
    \label{fig:ac-collecting}
\end{figure}

The basic assumption of this study was that (incorrect) answers given identically by a large number of students could represent an identical or similar error pattern and thus provide justification for further investigation. \autoref{fig:ac-collecting} shows the results of the frequency analysis.  In addition to the group of correct answers (first line, $AC_3$), we found many other relevant clusters of answers representing potential ACs. The largest of these groups were selected for further analysis, but only if they contained more than five answers. Limited resources were available for further evaluation, so we studied only the largest five groups per task in a group session with experienced educators to identify possible causes of the errors.

The groups for which objectively observable criteria could be found were included as new answer classes (see \autoref{fig:second-ac} for an excerpt of these). 
For example, although the task description clearly states that the start node should not be included in the list of visited nodes, some of the submissions showed it as the first node visited ($\text{AC}_{10}$).
Also, some of the nodes already visited were re-added to the list of visited nodes ($\text{AC}_{11}$).

While these examples are probably based on a careless reading of the task description, we also found ACs that indicate more profound technical errors.
Most ACs that emerged in the second iteration were much more focused on the causes of the answers given and made it possible to develop valuable feedback.
In task 4 (greedy search), there was a significantly large number of submissions (89 out of 462) that had the same error pattern: the detailed analysis revealed that this error pattern occurs when the greedy search algorithm is misapplied, namely by always selecting a neighbor of the last selected node instead of the overall cheapest node ($\text{AC}_{30}$). 
For the largest group of errors in $A^*$-search (52 of 462), the experts conjectured the cause to be the swapping of the edge direction between nodes $G$ and $I$. This could be due to the relatively small arrowheads in the task description, which may, in particular, pose a problem for students with impaired vision.

However, which misconceptions or knowledge gaps are underlying the respective AC cannot be determined with certainty without information from/about the learner. Further research would be necessary, for instance, with the help of interviews with the students. 

\begin{table*}[h]
\centering
\begin{tabular}[]{c | l | l}
    Task & ID & Answer Class\\
    \hline
    
    ST1-5 &$\text{AC}_{10}$ & start node included in the sequence of visited nodes\\
    ST1-5 &$\text{AC}_{11}$ & visited node included multiple times (circles/loops)\\
    ST1-5 &$\text{AC}_{12}$ & use of reverse alphabetical order to break ties \\
    ST1, ST2 &$\text{AC}_{21}$ & algorithm applied to heuristic instead of costs \\
    ST3, ST4 &$\text{AC}_{22}$ & algorithm applied to costs instead of heuristic \\
    ST2      &$\text{AC}_{31}$ & selection of nodes that lie in a horizontal line \\
    ST2      &$\text{AC}_{32}$ & selection of nodes that lie in a vertical line \\
    ST4      &$\text{AC}_{30}$ & greedy search considering only neighbors of previous node \\
    ST5      &$\text{AC}_{30}$ & A* search with reversed edge direction at nodes $G$ and $I$\\
\end{tabular}
\caption{Some answer classes newly found during the analysis}
\label{fig:second-ac}
\end{table*}

\section{Discussion}\label{sec:discussion}

Applying the steps described in Section \ref{sec:methodology} -- especially developing initial ACs -- required additional time. Whether the use of ACs saves time (see \textbf{ST} in \autoref{sec:potential}) cannot be answered based on the results in this paper. In the long term, however, we assume that time savings can be substantial, especially in shared tasks and settings. Even in task variants -- e.g., by ``changing numbers'' -- ACs can often be transferred (at least in spirit). In particular, if predominant error patterns are documented and suitable feedback has already been developed, the correction effort (and providing feedback) is thus reduced to the annotation with ACs. In our example of problem 2.1, many of the ACs found are general for tasks involving the application of search algorithms to graphs. 

While it is impossible to tell without further knowledge of the learner whether an answer ended up in a particular AC due to a misunderstanding or just sloppy work, it is possible to determine possible causes of errors and address them specifically in subsequent cohorts. Referring to $\text{AC}_{30}$ in ST4, we discussed in plenary what the reason could be that so many students incorrectly applied the greedy search algorithm. The instructor assumes that one prominent example given in the lecture was misread in a way that supported the false reading of the algorithm. In the example, a city network was used, which was traversed step by step. This can create the misconception to only pay attention to nodes directly connected to the previously visited node because, when physically traveling, it is usually not optimal to travel back to an already visited city to try out an alternative route. Also, the possible cause of $\text{AC}_{30}$ in ST5 made the educators decide to scale the arrows larger in the next exam.  

The use of answer classes can induce an improvement in feedback (\textbf{BF}), especially in cases where limited correction time disincentivizes writing the same (helpful) feedback into the answer sheets over and over again. Just marking the ACs and thus referring to a general feedback table for the exam can allow significant practical improvements of \textbf{BF}. Even if it is impossible to perform steps 4 and 5 for every answer class in large exams, it can still be used to handle the largest groups of answers. 

The question of whether ACs result in a more objective correction (\textbf{MOC}) can be answered clearly from our point of view. Due to predefined, objectively observable criteria, interpretations or subjective decisions are less likely, even when multiple humans are correcting an exam. If ACs are linked to numerical values, their combinations correspond to the task's scoring. It has also been found that corrections can be made much more fairly without additional effort. For instance, previously, students who started with the wrong starting node but performed the algorithm correctly did not get any points in the exam because all the mentioned nodes did not match the sample solution. However, these submissions are all in the same ACs (wrong starting node, correct algorithm). Thus, it is possible to realize a point deduction for the wrong start node but to consider consequential errors.   

The annotation of answers with ACs enables a higher quality evaluation of the exam results, which are of great use for the further development of teaching. Particularly pronounced error classes can indicate weaknesses in the teaching material that trigger misconceptions and may even suggest ways to heal them. Moreover, using suitable tasks in subsequent years may help verify whether the suspected causes and remedies were effective. Incorrect or ambiguous tasks could be identified and corrected in advance (STEP 1 and 2) (\textbf{BPE}), and potential for improvement could be identified after analyzing the submissions (STEP 3). Thus, applying the concept of answer classes stimulated incentives to improve teaching \textbf{IIT}, as the instructors developed new instructional materials that targeted uncovered problems and revised existing materials to avoid potential misconceptions in future courses.

\section{Beyond the Pilot}
\label{sec:vollkorn}

The results of the pilot show that a paper-based assignment of answer classes to learners' answers is possible and holds much potential to improve teaching processes. However, it is not particularly convenient to do it paper-based, and the potential of digitizing the process is obvious. As a result of the successful pilot, a system was developed with which answer classes can be assigned via a web interface to answers on paper-based exams -- the VoLl-KOrN system:
\VOLLKORN uses QR codes on the exam sheets containing relevant identifiers so learners' written answers can be automatically mapped to the corresponding task and person after scanning the exams. The advantage of this approach is that the sheets can be digitized in any order, and the risk of mixing up sheets after they have been stapled apart is no longer relevant. Using annotations in the source files of the exams, QR codes can be created automatically via a plugin. 

\begin{figure}\centering
    \includegraphics[width=\linewidth]{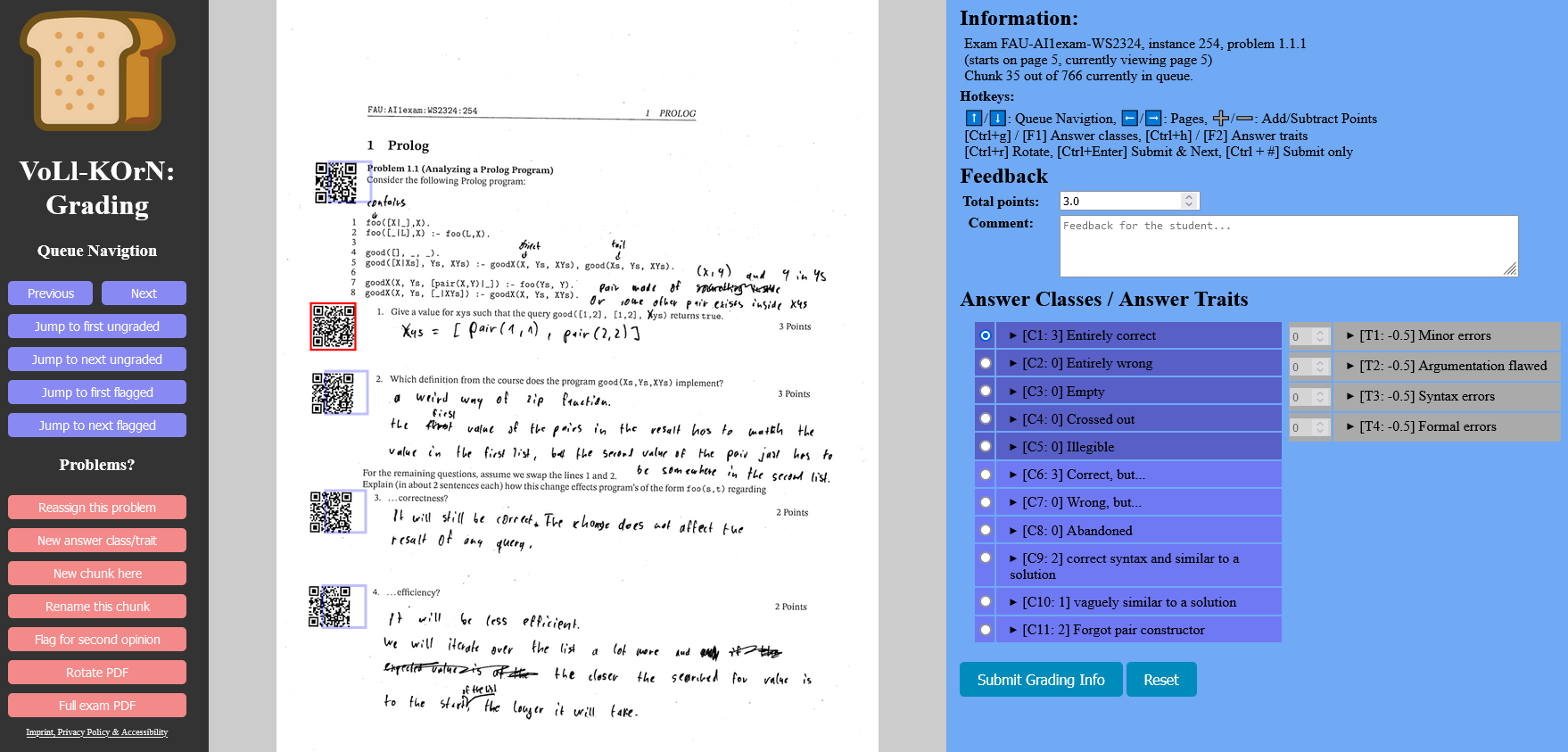}
    \caption{Assigning answer classes digitally with the VoLl-KOrN System}\label{fig:vollkorn}
\end{figure} 

The digitalization of the process opens up a range of benefits and new potential:
Graders can work collaboratively on correcting the exams from anywhere in the world and have the option of adding comments that are visible either as feedback for the examinees or the other correctors (\textbf{BF}). Linking answer classes with grading points also enables automatic evaluation of exam results based on answer classes. This saves time (\textbf{MOC}) and is also less prone to errors (\textbf{MOC}). 
In the context of adaptive learning systems, the information from the correction process can be processed directly back into the system and, for example, update corresponding learner models. This is particularly valuable when using a system like \VOLLKORN to grade weekly homework assignments. It is also possible to create cohort overviews (as is already established in systems that rely on learning analytics), which are a valuable source of information for tutors and instructors to address common errors and problems in subsequent sessions (\textbf{IIT}).

\section{Conclusion and Future Work}
\label{sec:conclusion-futurework}

We have picked up the concept of answer classes from the Y-Model framework \cite{LohBerKoh:ymodel23} and evaluated it in practice in a large-scale written exam. We have fleshed out a practical procedure for implementing answer classes in a written presence exam with manual paper-based correction. The results of this experiment are very encouraging for the quality of education by implementing answer classes. To quote the instructor of the course (who is a co-author of this paper): 

\begin{quote}\em
    \setlength{\leftmargin}{1em}
    \setlength{\rightmargin}{1em}
  \small{I was very skeptical whether this concept from secondary education could carry over to the university setting and whether the effort of AC annotation would be sustainable in a correction process that keeps my entire research group busy full-time for a week. Frankly, I only agreed to this to make my didactics colleague happy. [\ldots] But the result of the experiment has utterly convinced me of the approach, even in a paper-based setting. Seeing the discussions among the graders induced by developing and annotating ACs and the content awareness in the grading process made the added effort worthwhile. \quad [\ldots], and that is before we have taken a closer look at the data that this experiment generated. } 
\end{quote}

Encouraged by this, we digitized the process presented in Section \ref{sec:methodology} in a browser-based application that presents scanned exam papers and supports AC assignment, AC-based grading, AC-based feedback, and exam review also to reap the scalability benefits conjectured in Section \ref{sec:potential} and discussed in Section \ref{sec:discussion}. 

Developing and annotating ACs is time-consuming, and creating them individually in all situations may not be feasible. Therefore, future efforts should investigate the generalizability of answer classes, e.g., for similar and variant tasks, and whether any global answer classes can be used across all types of tasks.

We have not yet studied AC-based feedback generation and communication in detail. 
However, in the past, students received no feedback on their exam answers except from verbal discussions during exam review sessions.
Therefore, the limited experience from the experiment reported in this paper shows the potential of ACs to give standardized feedback that vastly improves the status quo. Future research must clarify whether AC-based feedback can keep up with individual feedback when including a learner model in the automated process. 

Finally, the automation of AC assignment is out of the scope of this paper and was left to future research.
In single/multiple-choice tasks the respective (combinations of) choices directly correspond to ACs, so automation is trivial.
However, about half of the exam tasks were more open-ended.
Already, for fill-in-the-blanks tasks, we have to pre-meditate or collect specific ACs, but given those, grading automation should be relatively easy to implement in general. For open-answer tasks, the problem of automating grading seems AI-hard -- i.e., if we can solve that, we can also solve the general AI problem. In situations where we have a lot more data -- with ACs and human classifications as a gold standard -- supervised deep learning methods might be successful in the future.

\newpage
\printbibliography
\end{document}